\documentclass[aps,prl,10pt,superscriptaddress,twocolumn,floatfix,longbibliography]{revtex4-1}
\usepackage[utf8]{inputenc}
\usepackage{amsmath}
\usepackage{amsfonts}
\usepackage{appendix}
\usepackage{amssymb}
\usepackage{graphicx}
\usepackage{hyperref}
\usepackage{listings}

\usepackage{xcolor}
\definecolor{DarkGreen}{rgb}{0.1,0.5,0.1}
\definecolor{DarkRed}{rgb}{0.5,0.1,0.1}
\definecolor{DarkBlue}{rgb}{0.1,0.1,0.5}
\hypersetup{
    unicode=false,         
    pdftoolbar=true,       
    pdfmenubar=true,       
    pdffitwindow=false,     
    pdfnewwindow=true,     
    colorlinks=true,      
    linkcolor=DarkBlue,         
    citecolor=DarkGreen,       
    filecolor=DarkGreen,     
    urlcolor=DarkBlue,         
    pdftitle={},
    pdfauthor={},
}
\usepackage{tikz}
\usetikzlibrary{positioning,shapes,arrows,fit}

\begin{document}
\title{Efficient near-optimal decoding of the surface code through ensembling}

\author{Noah Shutty}
\affiliation{Google Quantum AI, Venice, California 90291, USA}

\author{Michael Newman}
\affiliation{Google Quantum AI, Mountain View, California 94043, USA}

\author{Benjamin Villalonga}
\affiliation{Google Quantum AI, Venice, California 90291, USA}

\begin{abstract}
We introduce {\it harmonization}, an ensembling method that combines several ``noisy" decoders to generate highly accurate decoding predictions.
Harmonized ensembles of MWPM-based decoders achieve lower logical error rates than their individual counterparts on repetition and surface code benchmarks, approaching maximum-likelihood accuracy at large ensemble sizes.
We can use the degree of consensus among the ensemble as a confidence measure for a layered decoding scheme, in which a small ensemble flags high-risk cases to be checked by a larger, more accurate ensemble. 
This layered scheme can realize the accuracy improvements of large ensembles with a relatively small constant factor of computational overhead.
We conclude that harmonization provides a viable path towards highly accurate real-time decoding.
\end{abstract}

\maketitle

\section{Introduction}
A fast and accurate decoder is integral to the construction of a fault-tolerant quantum computer. It is responsible for continuously processing syndrome information from the quantum device to give corrected logical outcomes.
There are several efficient (i.e. polynomial-time) decoding algorithms \cite{demarti2023decoding}. Among the most promising are MWPM-based \cite{higgott2023sparse, higgott2023improved, fowler2013optimal} and union-find methods \cite{delfosse2022toward, huang2020fault} due to their speed.
On the other hand, there are more accurate but less-efficient decoders, such as maximum-likelihood (optimal) decoding with tensor networks \cite{bravyi2014efficient, chubb2021statistical, google2023suppressing, piveteau2023tensor} or most-likely-error decoding with integer programs \cite{feldman2005using,livontobel,fawzihypergraph,landahl2011fault}, whose runtime can scale exponentially with the decoding problem size.

Ideally, we would like a decoder that is both highly efficient and nearly optimal in accuracy.
Towards this end we introduce \emph{Harmony}, an assembly of correlated minimum-weight perfect matching decoders, each instantiated with a perturbed prior.
Individually, these decoders perform worse than a single matching decoder instantiated with a perfect prior.
Together, they form ensembled decisions that can approach maximum-likelihood inference.
We note that ensembling has also been used to boost the performance of decoding with machine learning \cite{sheth2020neural,bausch2023learning}.

This technique is embarrassingly parallelizable, and in the regimes we test, converges with reasonable ensemble sizes.
Querying the full set of decoders can lead to a large constant overhead, and so we introduce a layered decoding scheme to mitigate this cost.
In the layered scheme, a small group of watchful sentinels flag high-risk cases to be considered by the wider assembly.
The result is a highly accurate decoder with moments of expensive compute, but overall relatively low amortized cost.

First, we describe the correlated matching decoding algorithm that we use to build our ensembles.
Next, we describe Harmony, our ensembled correlated matching decoder.
We present our numerical results for the surface and repetition codes, comparing logical error rates of Harmony against a converged tensor network maximum-likelihood (TNML) decoder \cite{bohdanowicz2022quantum}.
For the repetition code, we optimize for 2D tensor networks using the sweep line method \cite{chubb2021general}.
For the surface code, we use a TNML decoder developed for the experiments of \cite{google2023suppressing}, and described in the Appendix.
Finally, we analyze the accuracy and cost of a layered decoding scheme, achieving the accuracy of a large ensemble with the amortized compute cost of a small ensemble.

\section{Correlated Matching}\label{sec:corrmatch}
In principle, any decoder could be used to build an ensemble, as long as there is a way of perturbing the error model prior.
In this work, we build our ensembles out of a correlated matching decoder that accepts as its prior an error hypergraph (or equivalently, a Tanner graph).
This decoder is similar to those described in
\cite{higgott2023improved,fowler2013optimal}.

Recall that an error hypergraph is a hypergraph for which each hyperedge is an error mechanism, each detector \cite{stim} is a vertex, and a hyperedge connects those detectors that the corresponding error activates.
Each hyperedge is labeled with the probability of the corresponding error mechanism and the set of observables it flips.
A minimum-weight perfect matching decoder can provide the most likely error hypothesis for an error \emph{graph}, on which each error mechanism activates at most two detectors.

In the surface code, the most common example of error mechanisms producing hyperedges with more than two detectors are $Y$-type errors.
We can decode the surface code by regarding $Y$-type errors as independent $X$-type and $Z$-type errors, where each error type forms a distinct, independent error graph.
However, this approximation neglects useful information that can assist decoding \cite{dennis2002topological}. 
A correlated matching decoder tries to reincorporate this information by reweighting the $X$- and $Z$-type error graphs as a function of the candidate matched edges in the opposite graph.

The correlated decoder proceeds in three steps: an initial minimum-weight perfect matching, a reweighting of edges based on that initial matching, and a final minimum-weight perfect matching.
In the initial minimum-weight perfect matching, we distribute the probabilities of edge-like error mechanisms to the corresponding $X$- and $Z$-type error graphs, ignoring the hyperedges.
We perform minimum-weight perfect matching on these disjoint error graphs (with weights set to $-\ln(p)$) to obtain a set of matched edges.
In the reweighting step, we assert that the matched edges have erred with certainty, and reweight other edges conditioned on this assertion.
This requires that we update the probabilities of those edges that together with the matched edge form hyperedges. 
We renormalize the probabilities of those hyperedges conditioned on the matched edge having erred and reassign these updated probabilities to the complementary edge in each hyperedge.
Over all such assignments (including the initial probability of the complementary edge) we select the one with the highest probability.
Finally, we run minimum-weight perfect matching on this reweighted graph to determine the correction.
In practice, we use Stim's \texttt{DetectorErrorModel} to configure the decoder \cite{stim}.

\begin{figure}
    \centering
    \includegraphics[width=\linewidth]{"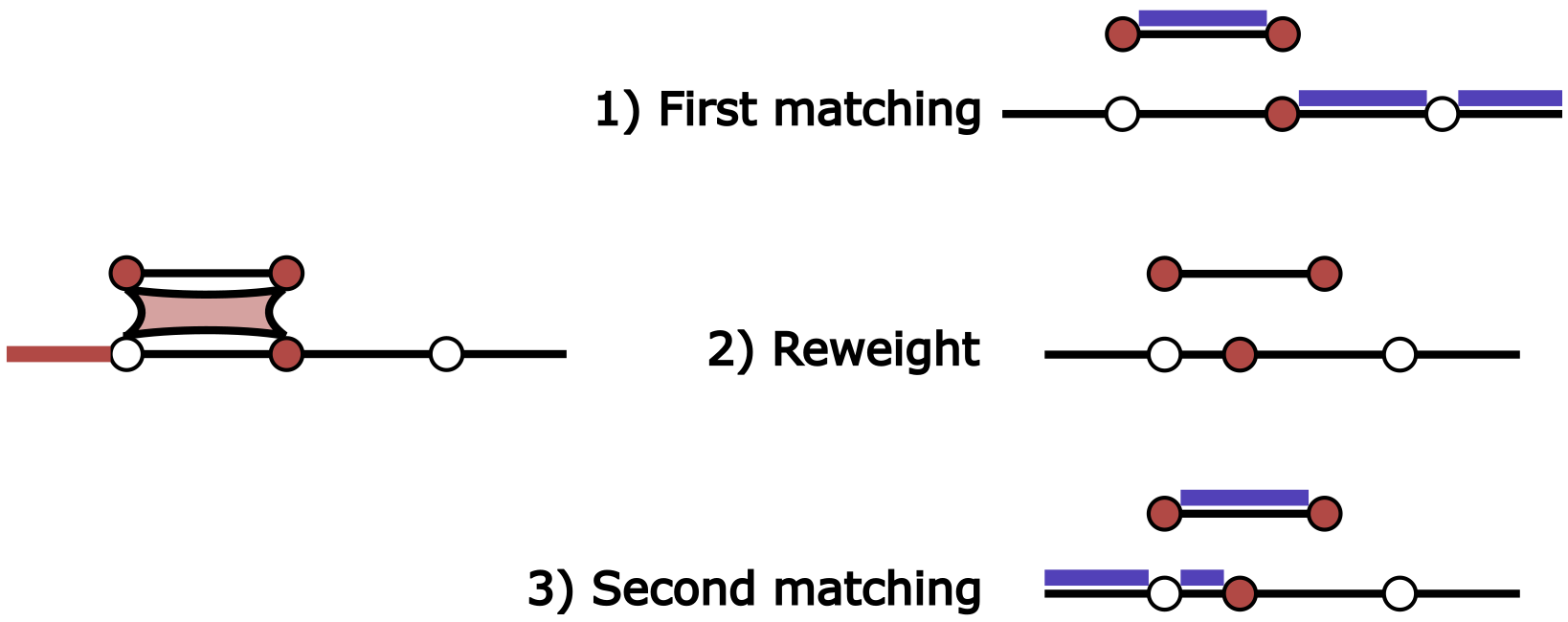"}
    \caption{
    A toy correlated matching example. 
    The error hypergraph is on the left, with the erring hyperedges and detection events shaded red, and with all edges having equal weight.
    First, we perform a minimum-weight perfect matching (blue edges) that misidentifies the error, in this case guessing the wrong of two equally valid choices.
    Second, we use the matching from the first step to infer that the hyperedge error mechanism has likely activated, and shorten the complementary edge.
    Third, we perform minimum-weight perfect matching on this reweighted graph, correctly identifying the error.
    }
    \label{fig:correlated_matching}
\end{figure}

\section{Harmonization}\label{sec:harmonization}
Conventionally, a correlated matching decoder is a deterministic algorithm that is configured with a fixed error model.
To harmonize a correlated matching ensemble, we have to solve two problems: ensemble generation and pooling. 
Ensemble generation consists of perturbing the error model prior given to each decoder so that the predictions of individual decoders begin to occasionally disagree.
Pooling consists of combining together all the decoder outputs into one overall decoding prediction. 

\subsection{Ensemble Generation}\label{sec:ensembleGen}
As described, correlated matching accepts an error hypergraph prior and transforms it to a graph where each edge $e$ is labeled with a probability $p_e$ and a set of implied probabilities for some other edges $S(e) = \{(e', q_{e'|e}): e'\text{ has probability } q_{e'|e}\text{ conditioned on the use of }e\}$.
In the ensembles described here, we directly perturb both the probabilities and implied probabilities of each edge. 
We produce a randomized graph where the edges are instead labeled with two perturbed probabilities $p^{(1)}_{e}, p^{(2)}_{e}$, where superscripts refer to the first and second passes of matching, and a perturbed reweight set $\tilde{S}(e) = \{(e', \tilde{q}_{e'|e}): e'\text{ has probability } \tilde{q}_{e'|e}\text{ conditioned on the use of }e\}$.

We modify correlated matching as follows. In the first pass, we use the edge probabilities $p^{(1)}_{e}$. Conditioned on the edges chosen in the first pass matching, we reweight some of the edges in the graph to have new probabilities as prescribed by the $\tilde{S}(e)$.
As before, we choose the maximum probability value for an edge that is reweighted multiple times.
We set the weights of any un-reweighted edge $e$ to $p^{(2)}_e$. Finally, we compute a second pass matching, which is the final decoding of the randomized ensemble member.

Now we must describe the probability distribution from which we sample $p^{(1)}_e, p^{(2)}_e, $ and $\tilde{S}(e)$.
We fix three constants $\alpha_1, \alpha_2, \alpha_3\in [0, 1]$. We sample the probabilities from the following intervals uniformly at random:
\begin{align}
    p^{(1)}_e &\sim \left[(1-\alpha_1)\cdot p_e, (1+\alpha_1)\cdot p_e\right]\\
p^{(2)}_e &\sim \left[(1-\alpha_2)\cdot p_e, (1+\alpha_2)\cdot p_e\right]\\
\tilde{q_{e'|e}} &\sim  \left[(1-\alpha_3)\cdot q_{e'|e}, (1+\alpha_3)\cdot q_{e'|e}\right].
\end{align}
Note that the set of edges reweighted by any given edge does not change, only the reweighted probability values.
The values of the constants $\alpha_i$ we use are:
$$\alpha_1 = 1, \,\,\, \alpha_2 = 4/5, \,\,\, \alpha_3 = 1/2.$$
Note that the above values were tuned for correlated matching on the surface code. Other codes and decoders may perform better with different perturbations.

\subsection{Pooling Methods}\label{sec:pooling}
\begin{figure}
    \resizebox{\linewidth}{!}{    
    \includegraphics{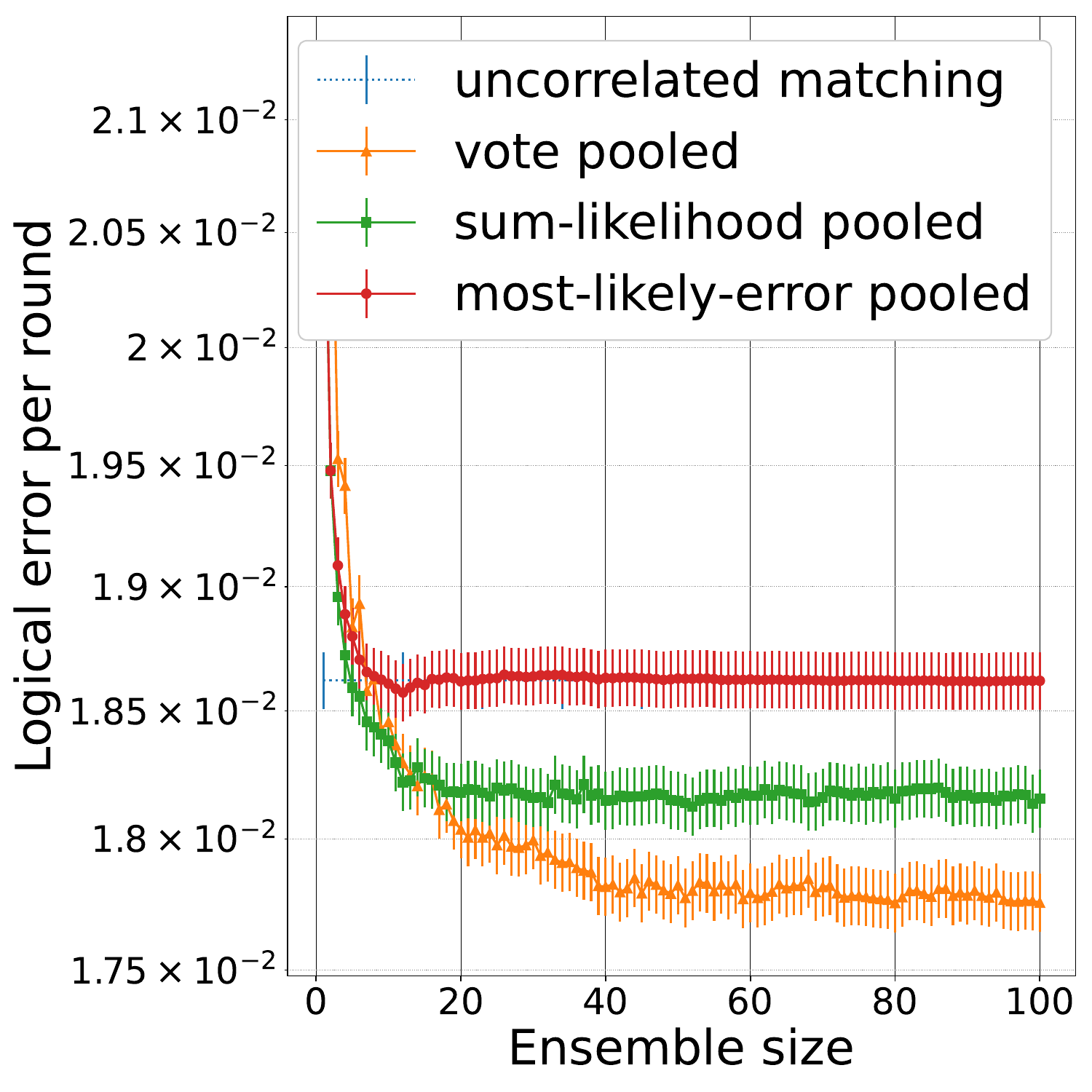}
    \includegraphics{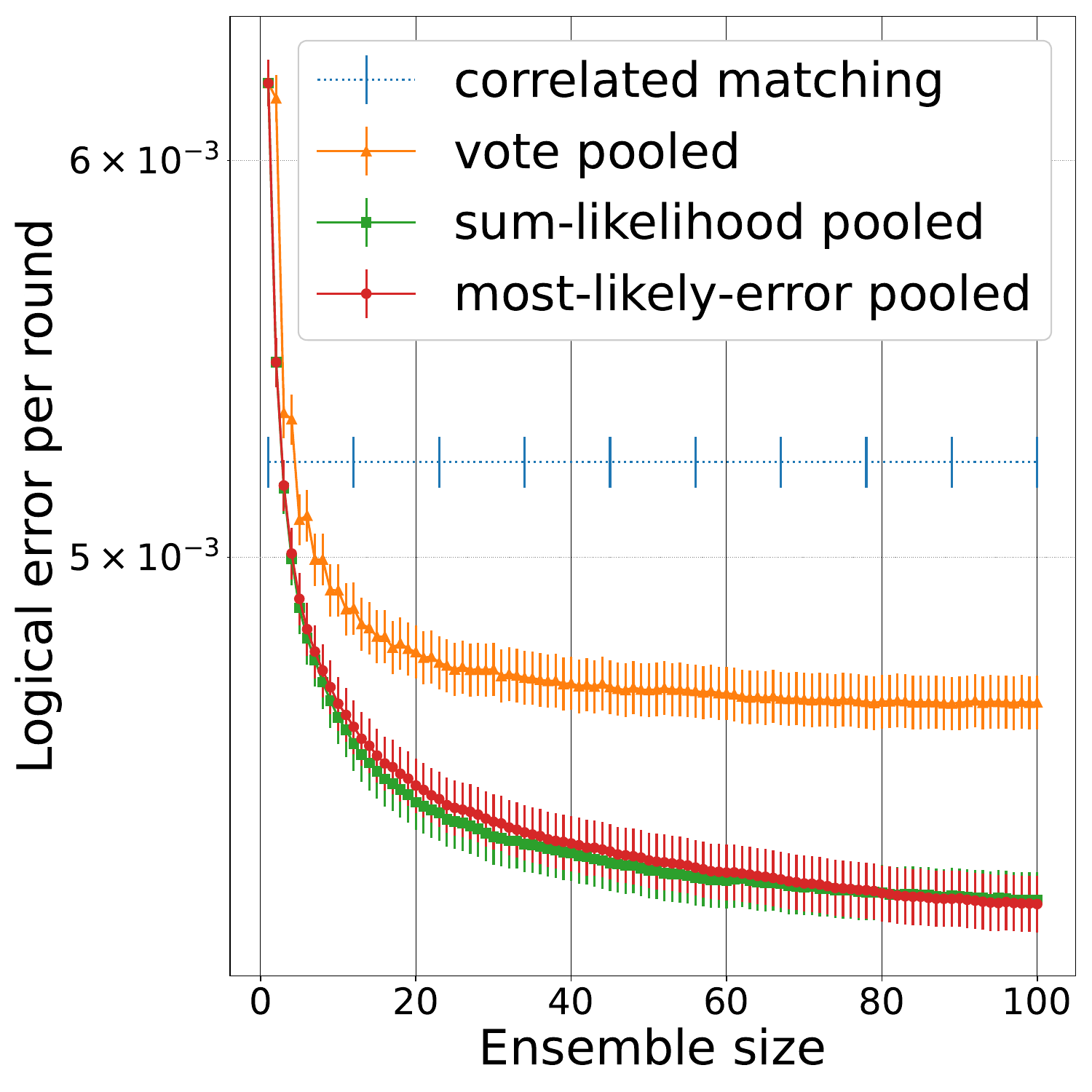}
    }
    \caption{Different pooling methods at distance-7 for circuit-level noise in the repetition code at $p=0.05$ (left) and the surface code at $p=0.008$ (right).
    In the repetition code, minimum-weight perfect matching is a most-likely error decoder, which most-likely-error pooling quickly converges to.
    We observe that vote pooling, which takes into account degeneracy, is the most accurate pooling method.
    For the surface code, most-likely-error pooling appears to be at least as effective as the other pooling methods once the ensemble is sufficiently large.
    This suggests that degeneracy plays a comparatively small role in increasing accuracy within the surface code.
    }\label{fig:poolingMethods}
\end{figure}
We consider three different pooling methods.
{\it Vote pooling} is perhaps the simplest and most directly generalizable pooling method. 
In this method we choose the decoding prediction given by the largest number of decoders.

Rather than giving each decoder an equal say in the pooling process, we may increase the influence of decoders that found a configuration of errors of relatively higher likelihood.
To make use of likelihood in pooling, each decoder must return the set of errors it used, rather than a set of edges.

\begin{figure}
    \centering
    \includegraphics[width=\linewidth]{"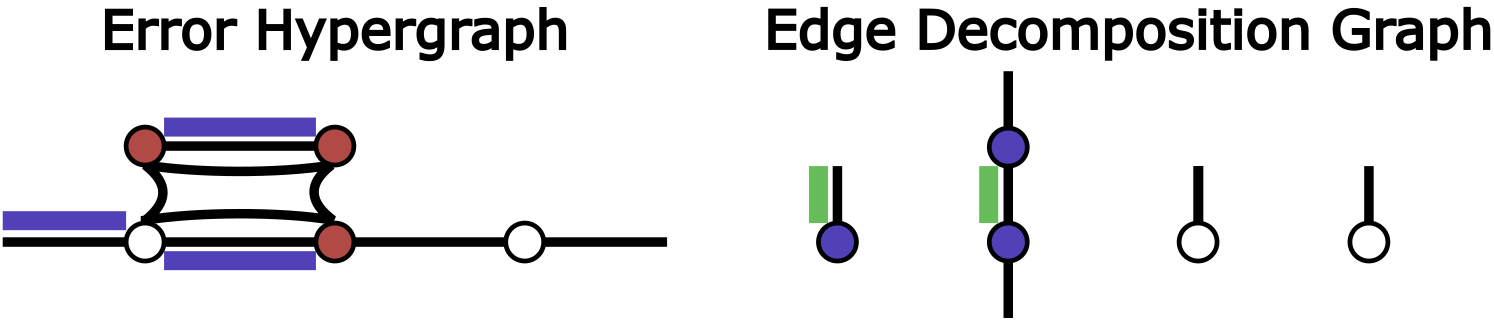"}
    \caption{
    On the left, an error hypergraph with a set of detection events shaded red and the edges chosen by a minimum-weight perfect matching decoder overlaid in blue.
    On the right, the corresponding edge decomposition graph, with nodes corresponding to chosen edges colored blue.
     MWPM-based decoding on the edge decomposition graph gives the most likely set of errors, overlaid in green, consistent with the edge set chosen by the matching decoder.
     In this case, it selects the far left edge-like error mechanism, as well as the only hyperedge-like error mechanism.
    }\label{fig:hedecomp}
\end{figure}

How can we recover the errors predicted during decoding (including those that induce hyperedges), given the edges used by a matching decoder?
We use an {\it edge decomposition graph} as described in Figure~\ref{fig:hedecomp}.
Vertices in this graph correspond to edge-like error mechanisms. 
Each edge connects two edge-like error mechanisms that combine to a hyperedge in the error hypergraph, while an edge-to-boundary at each vertex provides a way to match to nothing (representing an edge-like error).
In the conventional circuit noise model studied here, all independent error mechanisms cause at most one edge-like error in each of the $X$- and $Z$-type error graphs.
We can therefore apply Blossom to recover the most likely set of errors consistent with the chosen edges.

We consider two ways of pooling based on likelihood: {\it sum-likelihood pooling} and {\it most-likely-error pooling}. In sum-likelihood pooling, we sum the likelihoods of all decoders within each of the two possible logical flip predictions, and compare these sums to select the prediction with the largest value.
In most-likely-error pooling, we simply choose the decoder from the entire ensemble that reports the highest likelihood outcome and accept its prediction.

Generally, we observe that vote pooling gives the highest accuracy for the repetition code, while sum-likelihood and most-likely-error pooling give the highest accuracy for the surface code - see Figure~\ref{fig:poolingMethods}.

\section{Results}\label{sec:results}

To assess the accuracy of ensembled matching, we make three comparisons.
We use Stim's standard circuit generation tools to create decoding instances \cite{stim} - see the Appendix.

The first comparison we make is against maximum-likelihood decoding of repetition codes with circuit-level noise.
Famously, two-dimensional error graphs can be decoded optimally and efficiently using standard matrix product state techniques \cite{bravyi2014efficient}.
For circuit-level noise in the repetition code, we use the sweep line contractor developed in \cite{chubb2021general}.

\begin{figure}[ht]
    \centering
    \includegraphics[width=\linewidth]{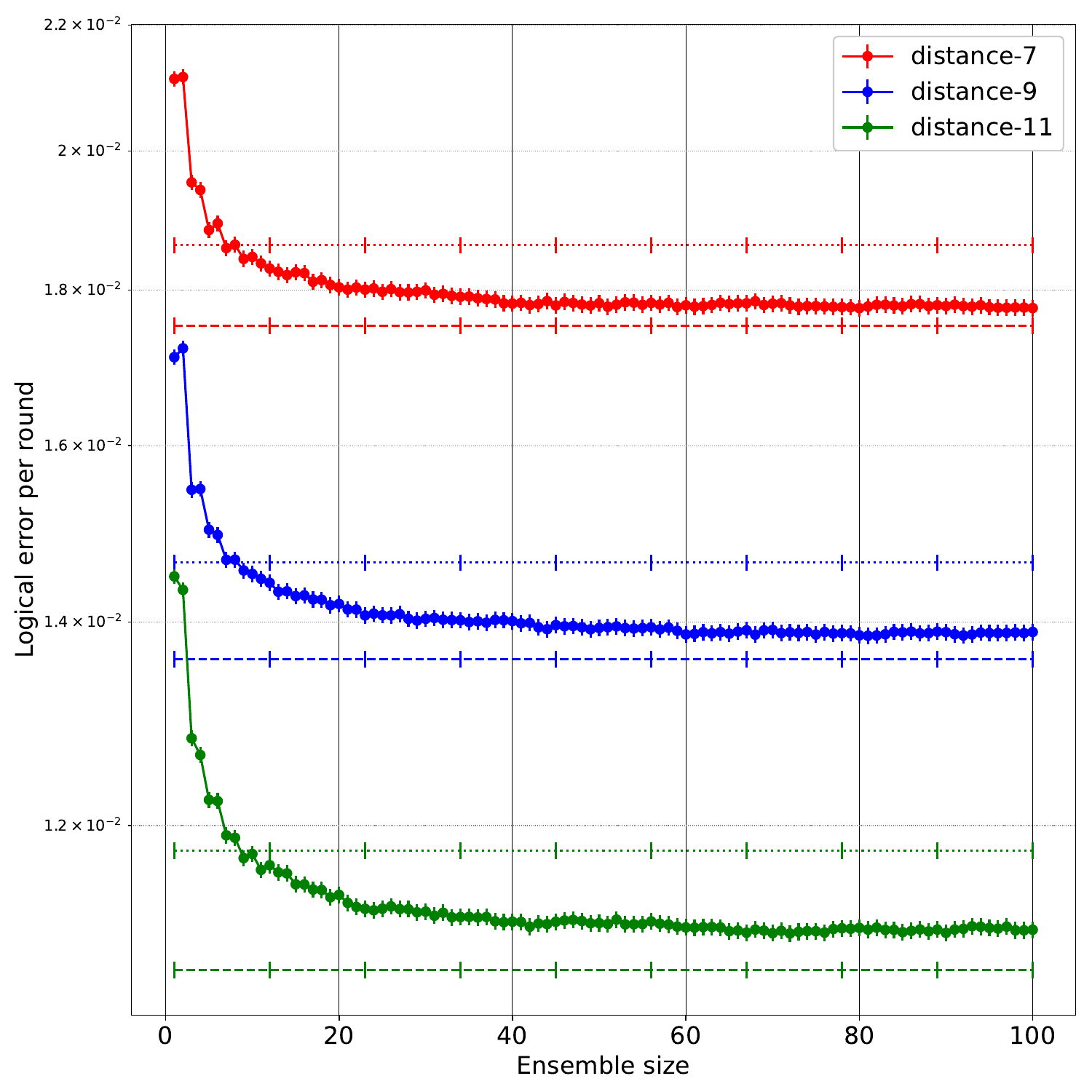}
    \caption{Decoding circuit-level noise in the repetition code at $p=0.05$. At each distance there are three lines: minimum-weight matching (dotted), vote pooled ensembled matching (circles), and TNML decoding (dashed). 
    We observe there is a relatively small difference between the most-likely error decoder and the TNML decoder, and this difference is mostly traversed at larger ensemble sizes.
    This is explained by the ensembled decoder accounting for degeneracy.
    }
    \label{rep_code_data}
\end{figure}

The repetition code is an interesting comparison because its decoding problem does not involve any hyperedges.
For this reason, minimum-weight perfect matching (when using $\ln(\frac{1-p}{p})$ edge weights) is an exact minimum-weight error decoder - only degeneracy within the error graph separates it from behaving as a maximum-likelihood decoder.
This allows us to isolate the ability of ensembling to account for degeneracy in decoding.
In principle, this effect also appears in the surface code, where the error graph of the repetition code embeds as a subgraph.
Note that most-likely-error pooling performs similar to minimum-weight perfect matching, as they both aim to return an error class with the lowest weight error.
Consequently, we elect to use vote pooling, which can account for degeneracy and in practice performs best - see Figure \ref{fig:poolingMethods}.
The results of this comparison are shown in Figure \ref{rep_code_data}, where we see that ensembling achieves near maximum-likelihood performance.

\begin{figure}[ht]
    \centering
    \includegraphics[width=\linewidth]{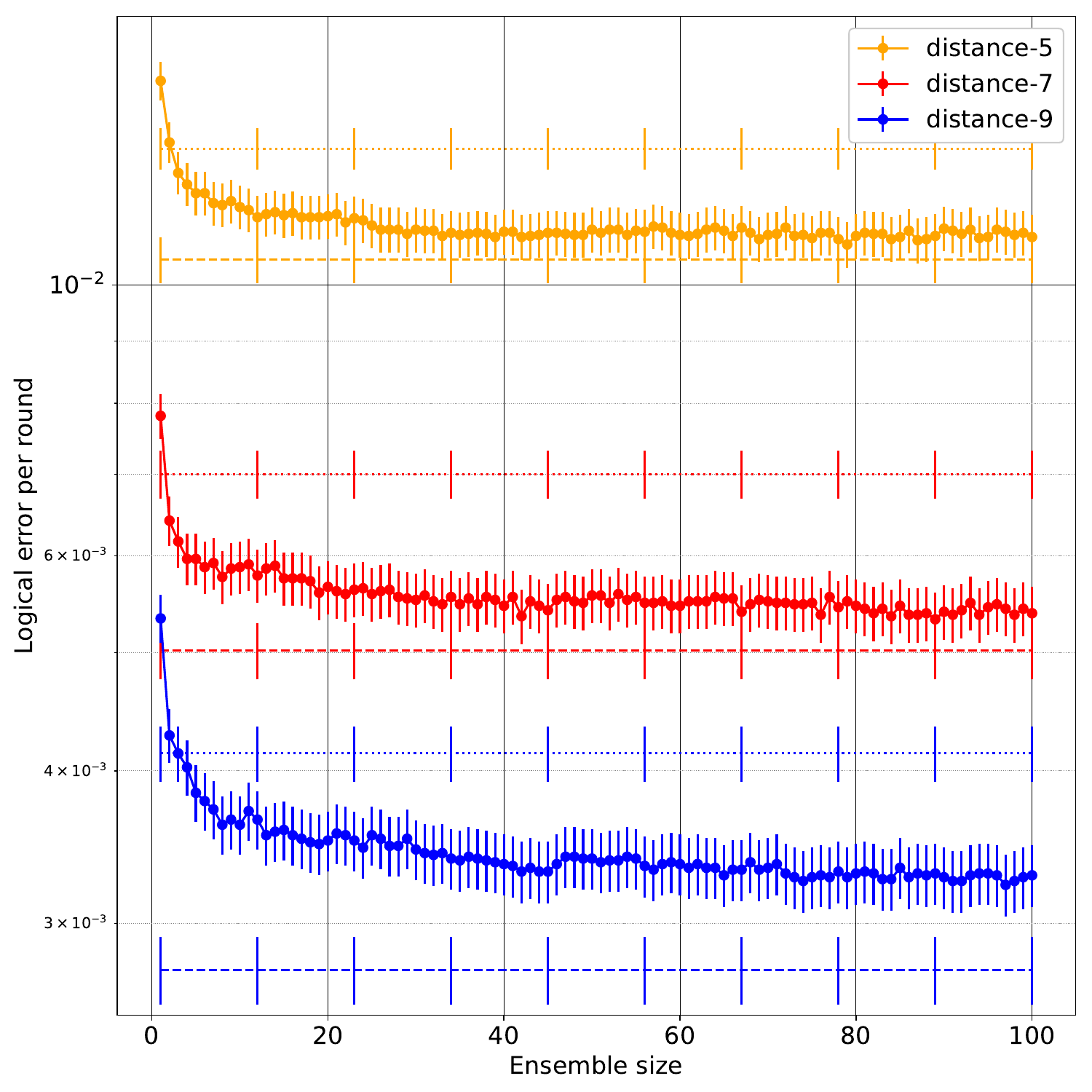}
    \caption{Decoding phenomenological noise in the surface code at $p=0.04$. 
    At each distance there are three lines: correlated matching (dotted), most-likely-error pooled ensembled matching (circles), and TNML (dashed).
    We observe that maximum-likelihood pooled MWPM is nearly as accurate as TNML once the ensemble size is large. As the distance increases, the gap between maximum-likelihood pooled MWPM and TNML increases slightly.}
    \label{phenom_surface_code}
\end{figure}

The second comparison we make is against a phenomenological noise model in the surface code.
We apply depolarizing noise to the data qubits (rather than the conventional bit-flip noise) to include hyperedges - see the Appendix for more details.
Because the error hypergraph of this problem is three-dimensional, TNML decoding is quite expensive \cite{piveteau2023tensor,google2023suppressing}.
To decode, we embed the decoding problem as a planar matrix product state contraction - also see the Appendix \cite{google2023suppressing}.
We chose phenomenological noise to reduce the size of these tensor networks so that contraction is not too computationally intensive at higher distances.
Interestingly, we note that most-likely-error and sum-likelihood pooling methods achieve similar performance - see Figure \ref{fig:poolingMethods}.
As most-likely-error pooling is aimed at finding the most likely error hypothesis, this suggests that in the surface code, most of the benefit of harmonization stems from better utilizing correlations rather than accounting for degeneracy.
We observe in Figure \ref{phenom_surface_code} that ensembled matching nearly bridges the accuracy gap between correlated matching and TNML decoding in this simplified error model.

\begin{figure}[ht]
    \centering
    \includegraphics[width=\linewidth]{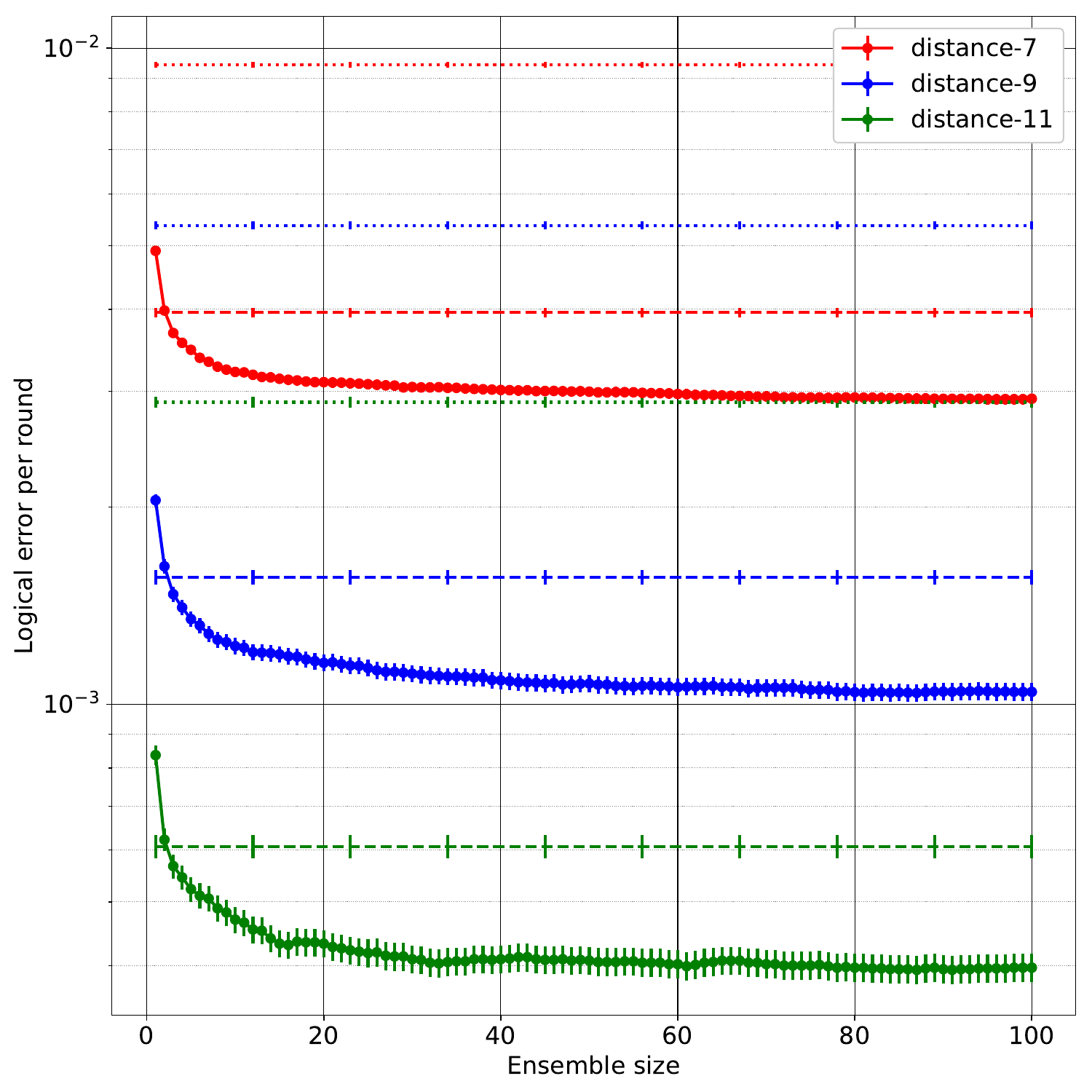}
    \caption{Decoding circuit-level noise in the surface code at $p=0.004$. 
    At each distance there are three lines: uncorrelated matching (dotted), correlated matching (dashed), and most-likely-error pooled ensembled matching (circles).
    The relative improvement grows with code distance.}
    \label{circuit_surface_code}
\end{figure}

Finally, having established that ensembled matching can achieve near maximum-likelihood performance, we study Harmony using full circuit-level noise in the surface code - see Figure \ref{circuit_surface_code} (and also the Appendix for additional data scanning over different error rates).
We find that a manageable ensemble size of 100 or so is more than sufficient to saturate most of the benefit achieved through ensembling, while an ensemble size of only 3 is sufficient to surpass correlated matching.
Of course, while ensembling is embarrassingly parallelizable, running 100 decoders in parallel would significantly increase the hardware cost of decoding.
However, a layered decoding scheme can reap most of the benefit while minimizing the overall cost.

\section{Confidence Rates and Layered Decoding}
\label{sec:layered}
Empirically, we observe that a low degree of consensus among decoders in the ensemble can suggest a particular decoding task is ``high-risk".
The fraction of decoders that vote with the majority can therefore be used as a confidence score for the decoding.
In a try-until-success protocol, e.g. in certain magic state preparation routines, we could simply restart whenever the confidence of decoding is unacceptably low.
For general computation, we cannot afford to start from scratch every time the ensemble has low confidence.
We therefore consider an alternative use for the confidence scores: layered decoding.
In a layered decoding scheme, a lighter-weight first-pass decoder layer handles all decoding for a majority of the shots, only passing ``difficult" shots where it is not sufficiently confident in its prediction to a slower but more accurate second-pass decoding layer. 

\begin{figure}
    \centering
    \includegraphics[width=1\linewidth]{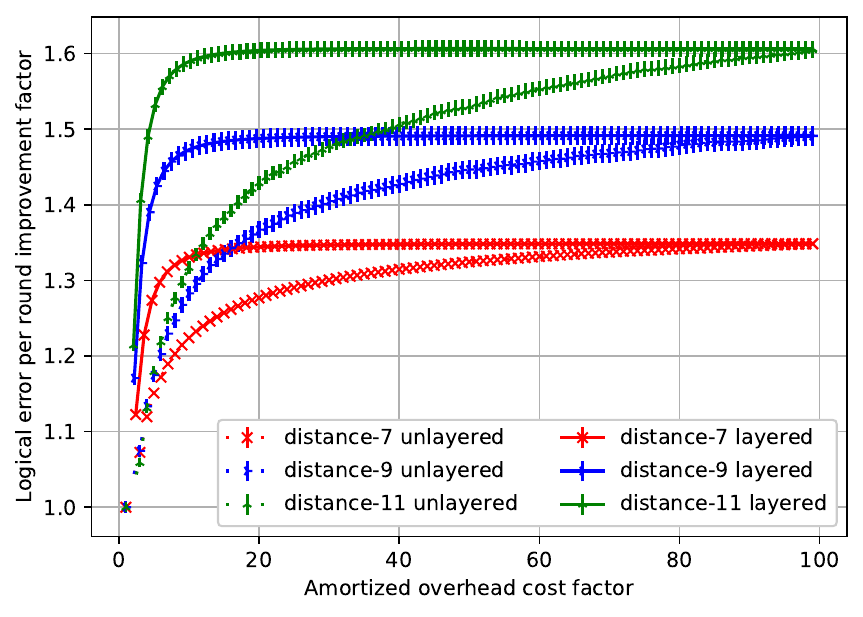}
    \caption{The improvement in logical error per round versus the overhead cost factor for a one-pass unlayered decoding scheme (dotted line) and a two-pass layered decoding scheme (solid line) for circuit-level noise in the surface code at $p=0.004$.
    The first pass is a Harmony ensemble of size $N_1$ (varying) and the second pass decoding is given by a most-likely-error pooled Harmony ensemble of size $N_2 = 100$. The second pass decoding is executed if and only if the first pass ensemble fails to yield a consensus prediction, which happens with some probability called the ``trigger rate". The horizontal axis is the expected number of decoding instances per shot, which increases with $N_1$ because both the first pass ensemble size is larger and because the trigger rate increases due to a higher probability of a dissenting member in the larger first pass ensemble.
    Layered decoding achieves the accuracy of a much larger ensemble without significant amortized overhead.
    }
    \label{fig:layeredDecodingScheme}
\end{figure}

We consider a simple two-pass scheme in which an ensemble of $N_1$ decoders is used in the first pass, and it activates the second pass only when the confidence is less than $100\%$ (i.e., if there is any dissent among the decoders). The improvement factor in logical error rate and the overhead are shown in Fig.~\ref{fig:layeredDecodingScheme}. The results show that we saturate almost all of the improvement in logical error rate with $N_1 \approx 4$.
For lower error rates, the amortized overhead cost per shot is very close to $N_1$, since the first pass ensemble almost always reaches a unanimous decision.
We note that one could also use the complementary gap \cite{gidney2023yoked} to flag high-risk cases in the first-pass. 
Plausibly, this could bring the amortized cost close to $N_1 \approx 1$, but we leave comparing these methods to future work.

\section{Conclusion}
We introduced harmonization as a simple and effective form of ensembling for decoders. 
We showed that harmonization gives near-optimal logical error rates for the repetition code and the surface code by benchmarking against TNML decoding.
The main improvement of ensembling comes from degeneracy for the case of the repetition code, and from better use of correlations for the case of the surface code.
Moreover, the degree of consensus among the ensemble gives a confidence score that can be exploited to significantly reduce the amortized overhead of ensembling in a layered decoding scheme.
This suggests that harmonization is a promising path to efficiently reduce the logical error rate of real-time decoders.
While we have focused on matching in surface codes, many exciting avenues remain, and we expect these techniques could greatly improve other decoders and codes.

\section{Acknowledgements}
\begin{acknowledgements}
We thank Sergio Boixo, Dripto Debroy, Austin Fowler, Craig Gidney, Cody Jones, and Adam Zalcman for helpful conversations and technical insights.
\end{acknowledgements}

\bibliography{refs.bib}

\appendix
\section{Generating Circuits}
\label{appendix:generating_circuits}

For reproducibility, we use Stim's \cite{stim} standard circuit generation features to create noisy circuits.
With circuit-level noise, these are similar but slightly different to `standard' circuit-level depolarizing models.
They aggregate measure and reset error into a single error, do not include idling error, and add a depolarizing channel before each cycle of error-correction.
These circuits can be reproduced using the following Stim commands. For circuit-level noise in the repetition code:
\begin{lstlisting}
stim gen \
--code repetition_code \
--task memory \
--distance $distance_var \
--rounds $rounds_var \
--before_round_data_depolarization $p_var \
--before_measure_flip_probability $p_var \
--after_reset_flip_probability $p_var \
--after_clifford_depolarization $p_var
\end{lstlisting}
For phenomenological noise in the surface code:
\begin{lstlisting}
stim gen \
--code surface_code \
--task rotated_memory_z \
--distance $distance_var \
--rounds $rounds_var \
--before_round_data_depolarization $p_var \ 
--before_measure_flip_probability \
$(echo "2*$p_var/3" | bc -l)
\end{lstlisting}
Note the $2p/3$ probability for the measurement error.
This is chosen so that each individual $X$- or $Z$-type error graph has uniform edge-weights when treating $Y$-type errors as independent $X$- or $Z$-type errors. This matches the usual (uncorrelated) phenomenological decoding of a cubic error graph. Finally, for circuit-level noise in the surface code:
\begin{lstlisting}
stim gen \
--code surface_code \
--task rotated_memory_z \
--distance $distance_var \
--rounds $rounds_var \
--before_round_data_depolarization $p_var \
--before_measure_flip_probability $p_var \
--after_reset_flip_probability $p_var \
--after_clifford_depolarization $p_var
\end{lstlisting}

Note that simulations for the repetition code and circuit-level noise in the surface code use $2d$ rounds. For phenomenological noise in the surface code, we use $d$ rounds to keep the simulations more manageable for TNML decoding.

\section{Maximum-likelihood decoding with tensor networks}\label{appendix:tensorNetworkDecoder}

We benchmark the results shown in the main text against an (approximate) maximum-likelihood decoder.
There are several proposed implementations for a circuit-level tensor network decoder, including maximum-likelihood decoding of the circuit history code \cite{bohdanowicz2022quantum}.
However, we instead use the tensor network maximum-likelihood (TNML) decoder developed for the experiments of \cite{google2023suppressing}.
This decoder accepts as input any error model that can be expressed as an error hypergraph, as well as a list of detection events, and outputs the likelihood that the local observable either changed parity or not conditioned on the observed detection events. 
In this Appendix we describe our TNML decoder.

\subsection{The error hypergraph}
\label{sec:error_hypergraph}

An error hypergraph can be represented as a bipartite (Tanner) graph where one subset of nodes ($A$) represents the error mechanisms contained in the model and the other subset of nodes ($B$) represents the union of all detectors and logical operators.
For each error mechanism, of independent probability $p_i$, we have a binary variable $e_i$ that denotes whether the error is present (1) or not (0).
Similarly, for each detector we have a binary variable $d_j$ denoting whether the detection event is present or not.
Finally, during a memory experiment each logical operator might change value (flip), a process that is also represented by a binary variable $l_k$.
As an example, consider the following error hypergraph, arising from a distance-3 surface code over a single round and with a single logical observable:
\begin{align}
    \label{eq:error_hypergraph}
    \vcenter{\hbox{\includegraphics[width=0.50\linewidth]{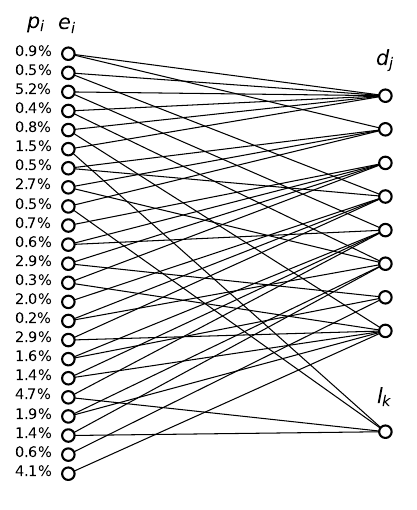}}}
\end{align}
The connectivity of the error hypergraph represents the relation between an error configuration (bit string over $A$) and the detection event and logical operator configuration (bit string over $B$): given an error configuration over $A$, the value of a variable on $B$ is given by the parity of the variables on $A$ incident on it.
In other words, both the detector configuration $\vec{d}$ and the logical observable configuration $\vec{l}$ are functions of the error configuration: $\vec{d} = \vec{f}(\vec{e})$ and $\vec{l} = \vec{g}(\vec{e})$.
The following is an example error configuration over the error hypergraph of \eqref{eq:error_hypergraph}:
\begin{align}
    \label{eq:error_conf}
    \vcenter{\hbox{\includegraphics[width=0.50\linewidth]{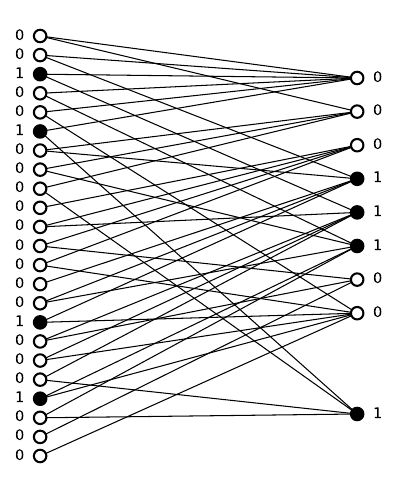}}}
\end{align}

The probability of an error configuration $\vec{e} \equiv \left\{ e_i \right\}_i$ is equal to:
\begin{align}
\label{eq:probability_error_conf}
    {\rm Pr}(\vec{e}) = \prod_i p_i^{e_i} \cdot (1 - p_i)^{1 - e_i},
\end{align}
i.e. each error that is present contributes a factor $p_i$ and each error that is not present contributes a factor $(1 - p_i)$.

\subsection{A tensor network for decoding}
\label{sec:tensor_network_decoding}

The goal of a decoder is to infer which value of the logical observable(s) is most likely conditioned on a particular configuration of detection events (detector configuration) observed in an experiment.
We denote this likelihood by $L \left( \vec{l} | \vec{d} \right)$.
The most likely value of $\vec{l}$ is then chosen as the most plausible logical observable configuration.
Formally, the likelihood is computed as:
\begin{align}
\label{eq:likelihood}
    L \left( \vec{l} | \vec{d} \right) \propto \sum_{\vec{e} : \left[ \vec{f}(\vec{e}) = \vec{d} \right] \land \left[ \vec{g}(\vec{e}) = \vec{l} \right] } {\rm Pr}(\vec{e})
\end{align}
i.e. as the sum of the probabilities of all error configurations compatible with the observed detection event configuration and a specific logical observable configuration.
In the case of having a single logical observable $l_0$, the inferred value is equal to 1 if $L \left( l_0=0 | \vec{d} \right) \geq L \left( l_0=1 | \vec{d} \right)$ and 0 otherwise.

It is possible to evaluate the sum in Eq.~\eqref{eq:likelihood} by contracting a tensor network.
This tensor network has the same topology as the error hypergraph.
Each node corresponding to an error mechanism $e_i$ with associated probability $p_i$ can be regarded as a tensor of the form
\begin{align}
    \label{eq:e_tensor}
    \vcenter{\hbox{\includegraphics[width=0.30\linewidth]{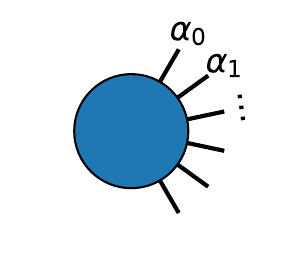}}}
    =
    \begin{cases}
    p_i &{\rm if} \,\, \alpha_0 = \alpha_1 = \ldots = 1 \\
    1 - p_i &{\rm if} \,\, \alpha_0 = \alpha_1 = \ldots = 0 \\
    0 \quad &{\rm otherwise}
    \end{cases}
\end{align}
i.e. a Kronecker delta (also known as a copy tensor) weighted with the values $p_i$ and $1-p_i$.
Each node corresponding to a detector $d_i$ can be regarded as a tensor of the form
\begin{align}
    \label{eq:d_tensor}
    & \vcenter{\hbox{\includegraphics[width=0.30\linewidth]{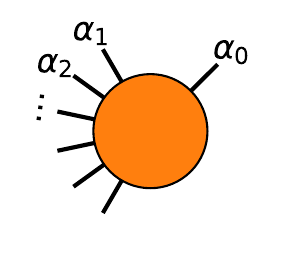}}}
    =
    \begin{cases}
    1 & {\rm if} \,\, \alpha_0 + \alpha_1 + \ldots \,\, {\rm even} \\
    0 & {\rm if} \,\, \alpha_0 + \alpha_1 + \ldots \,\, {\rm odd}
    \end{cases}
\end{align}
i.e. a tensor that enforces even parity over the union of the errors incident on the detector and the detection event variable itself, $d_i$. 
Each logical operator is also substituted by a tensor of the from of Eq.~\eqref{eq:d_tensor}, enforcing even parity over the errors incident to it and the logical operator variable $l_k$.
For example, the error hypergraph in \eqref{eq:error_hypergraph} results in the tensor network
\begin{align}
    \label{eq:tanner_tn}
    L \left(l_0 | \vec{d} \right) = 
    \vcenter{\hbox{\includegraphics[width=0.67\linewidth]{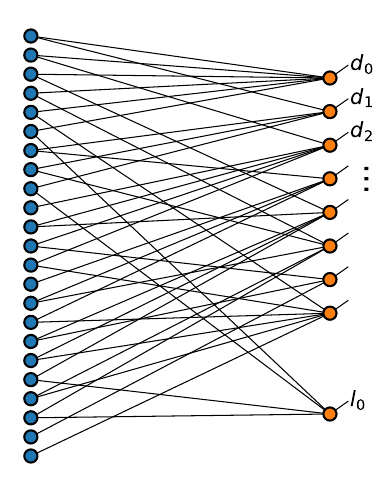}}}
\end{align}
Given a detection event configuration $\vec{d}$, one can project all indices $d_j$ to their corresponding value.
For each value of $l_0=0, 1$, the contraction of the tensor network yields a scalar equal to $L\left( l_0 | \vec{d} \right)$.
Alternatively, after projecting all $d_j$, one can leave the $l_0$ index open (not projected), in which case the contraction of the tensor network yields a vector with two entries, $L \left( l_0=0 | \vec{d} \right)$ and $L \left( l_0=1 | \vec{d} \right)$.
We can then compare both likelihoods to make a decoding decision.

Understanding why the contraction of the tensor of Eq.~\eqref{eq:tanner_tn} yields the claimed value is relatively straightforward.
To start with, the contraction of a tensor network is equal to the sum over all configurations of its indices (all binary in our case) of the product of the tensor entries corresponding to each index configuration.
Our tensor network has two types of tensors.
The first one corresponds to each error mechanism and its tensor entries yield 0 unless all indices incident on it are equal, resulting in only two contributing configurations.
If they are all equal to 1, then the error is considered present and the tensor contributes a factor $p_i$.
If they are all equal to 0, then the error is not present and the tensor contributes a factor $(1 - p_i)$.
The other type of tensors is parity tensors.
For each of these, if the incident detection event is present (1), then the remaining indices have to add up to odd parity, or else the tensor will contribute with a factor of 0.
If the incident detection event is not present (0), then the remaining indices have to have even parity, or else the tensor will contribute with a 0.
This imposes the constraint that the errors incident to each detector must be consistent with the presence or absence of a detection event.
This logic applies identically to the logical observables.
In turn, the contraction of the tensor network yields the sum of all products of factors $p_i$ (error present) and $(1 - p_i)$ (error not present) for all configurations of errors that are compatible with the observed detection event configuration and logical observables, exactly as in Eq.~\eqref{eq:likelihood}.

In practice, contracting the tensor network of Eq.~\eqref{eq:tanner_tn} can be challenging.
The computational cost of a tensor network contraction is dominated by a factor that is exponential in the tree width of its line graph, a property that is related to the topology of the tensor network, or the error hypergraph in our case.
For a surface code, the cost of the tensor network contraction scales exponentially in $O({\rm min}(d^2, c\cdot dr))$, where $d$ is the distance of the surface code, $r$ is the number of rounds, and $c$ is some positive scalar (which roughly converts ``time'' or ``round'' units into spatial units).
This makes exact tensor network contraction decoding of distance $\geq 5$ surface codes impractical.
We have to resort to an approximate contraction of the tensor network which, when converged, is just as accurate.

\subsection{Contracting the tensor network in practice: matrix product states}

Given the intractability of contracting the tensor network $L\left( \vec{d} | \vec{l} \right)$, we rearrange it to a form in which it can be easily contracted approximately.
Our goal is to rewrite the tensor network of Eq.~\eqref{eq:tanner_tn} as a planar graph on a square grid, which can be contracted approximately as a matrix product state (MPS) evolution with maximum bond dimension $\chi$~\cite{vidal2003efficient}.

We first note that a parity tensor like the one in Eq.~\eqref{eq:d_tensor} can be expanded as an MPS of dimension 2:
\begin{align}
    \label{eq:parity_mps}
    \vcenter{\hbox{\includegraphics[width=0.27\linewidth]{d_tensor.pdf}}}
    =
    \vcenter{\hbox{\includegraphics[width=0.22\linewidth]{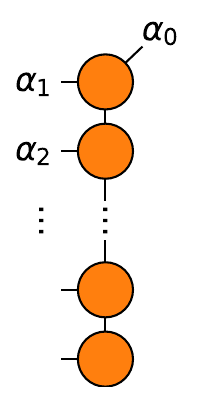}}}
\end{align}
where each tensor on the right is itself a parity tensor as defined in Eq.~\eqref{eq:d_tensor}.
With the equivalence of Eq.~\eqref{eq:parity_mps} in mind, we can rewrite the tensor network of Eq.~\eqref{eq:tanner_tn} with a trivial substitution of parity tensors.
We can further manipulate the resulting tensor network so that each tensor it is easily placed on a 2D grid.
In our example, that is:
\begin{align}
    \label{eq:2d_tn}
    L \left(l_0 | \vec{d} \right) = 
    \vcenter{\hbox{\includegraphics[width=0.67\linewidth]{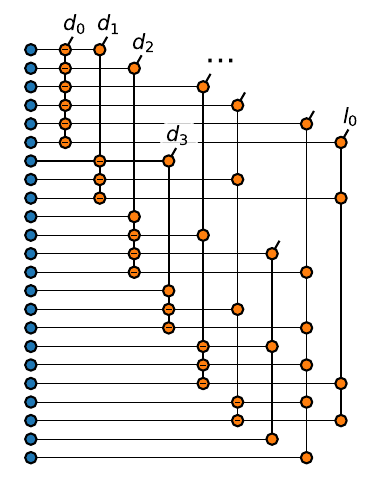}}}
\end{align}
where we use the modified parity tensors
\begin{align}
    \label{eq:modified_parity}
    & \vcenter{\hbox{\includegraphics[width=0.30\linewidth]{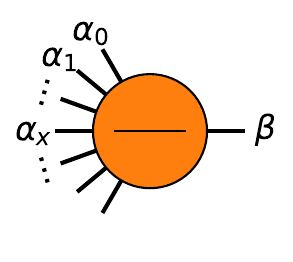}}}
    =
    \begin{cases}
    1 & {\rm if} \,\, (\alpha_0 + \alpha_1 + \ldots \,\, {\rm even}) \\
      & \,\,\,\,\,\, \land \,\, (\alpha_x = \beta) \\
    0 & {\rm otherwise}
    \end{cases}
\end{align}
i.e., a parity tensor over $\vec{\alpha}$ that also acts as a copy tensor between an index $\alpha_x$ and $\beta$.
This is useful so that each error mechanism index can ``propagate'' to the detector MPS applied further to its right.
This way, each error tensor ``propagates'' its only index indefinitely to the right and a single row is needed per error mechanism.
Note that the resulting 2D grid has as many rows as error mechanisms and as many colums as detectors.
In addition, parity tensors are placed according to the adjacency matrix of the bipartite Tanner graph, i.e. each coordinate of this grid has a parity tensor if and only if the error mechanism corresponding to that row and the detector corresponding to that column have an edge in common in the error hypergraph.

Finally, note the crossing indices in the tensor network of Eq.~\eqref{eq:2d_tn}.
In order to obtain a tensor network over a 2D grid that is also a planar graph, we add trivial crossing tensors on each intersection of indices:
\begin{align}
    \label{eq:crossing_tensor}
    & \vcenter{\hbox{\includegraphics[width=0.30\linewidth]{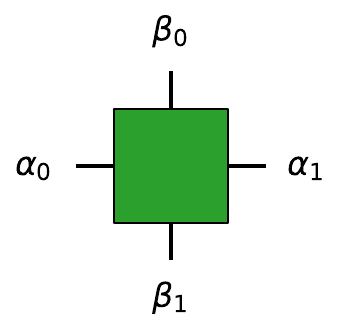}}}
    =
    \begin{cases}
    1 & {\rm if} \,\, (\alpha_0 = \alpha_1) \,\, \land \,\, (\beta_0 = \beta_1) \\
    0 & {\rm otherwise}
    \end{cases}
\end{align}
which ``propagate'' indices horizontally \emph{and} vertically.
In our example, this results in the final tensor network
\begin{align}
    \label{eq:plannar_tn}
    L \left(l_0 | \vec{d} \right) = 
    \vcenter{\hbox{\includegraphics[width=0.67\linewidth]{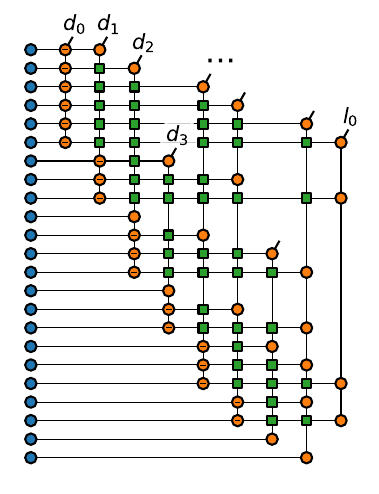}}}
\end{align}

The planar tensor network of Eq.~\eqref{eq:plannar_tn} can now be approximately contracted as the evolution of an MPS with maximum bond dimension $\chi$ from left to right.
The initial MPS is a product state where each site $i$ is in the $(p_i, 1-p_i)$ superposition.
Each detector MPS (technically a matrix product \emph{operator} or MPO on the indices that are propagated through the copied index) is applied to the evolving MPS.
When a site $i$ reaches the MPO or column where error $e_i$ reaches its last detector, then that site does not propagate and disappears from the evolution.

\subsection{Approximate contraction of the tensor network with maximum bond dimension \texorpdfstring{$\chi$}{chi}$\chi$}

\begin{figure}
    \centering
    \includegraphics[width=1.0\linewidth]{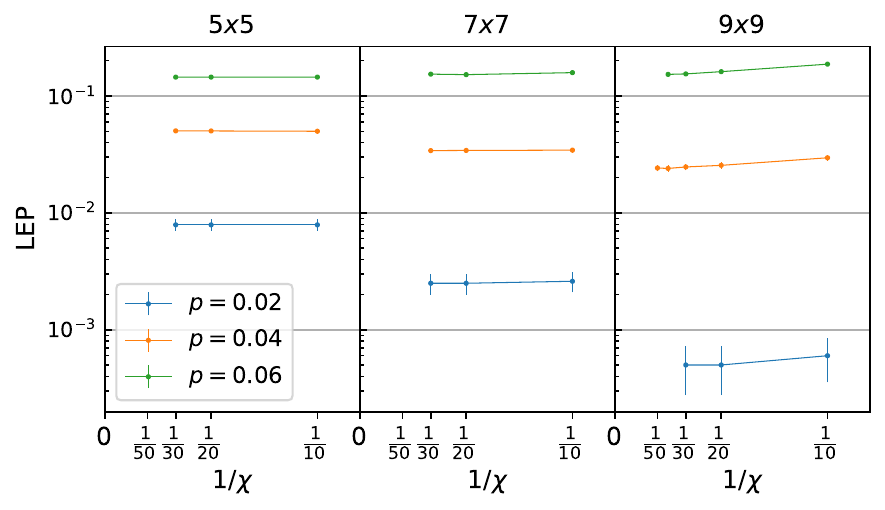}
    \caption{
    Convergence of the logical error probability (LEP) using the TNML decoder as a function of the bond dimension $\chi$.
    We show the results for surface codes of distance $d = 5, 7, 9$ over $r=d$ rounds.
    All results are converged up to statistical uncertainty over the 10,000 shots decoded.
    }
    \label{fig:ml_convergence}
\end{figure}

As explained above, the contraction of the tensor network of Eq.~\eqref{eq:plannar_tn} can be approximated by an MPS evolution of maximum bond dimension $\chi$.
Such contraction yields an approximate likelihood
\begin{equation}
\label{eq:approximate_L}
    L_\chi\left(\vec{d} | \vec{l}\right) \approx L\left(\vec{d} | \vec{l}\right) \,.
\end{equation}
We now analyze the convergence of the approximate contraction $L_\chi$ described in this section as a function of $\chi$.
In particular, we look at the final logical error probability (LEP) when decoding using approximate maximum-likelihood decoding with bond dimension $\chi$.
Fig.~\ref{fig:ml_convergence} shows the LEP as a function of $\chi$.

\subsection{Computational cost of the TNML decoder}
\label{sec:computational_cost}

In this section we restrict our analysis to surface codes.
The resulting planar tensor network resembles a circuit applied to a 1D register of qubits, albeit describing a non-unitary evolution.
The width of the circuit, equal to the number of error mechanisms in the error model, is proportional to the volume of the computation $d^2 r$.
For an error model that is local in time and both spatial dimensions, such as the ones considered in this paper and commonly used in practice, and for a ``good'' ordering of error mechanisms and detectors, the effective circuit depth for each site (error mechanism) is proportional to $d^2$.
This is because error mechanisms affect only a small number of rounds --- typically at most two in conventional circuit-level noise models.
See for example the following distance-5 surface code over 5 rounds:
\begin{align}
    \label{eq:d5_adjacency}
    \vcenter{\hbox{\includegraphics[width=0.30\linewidth]{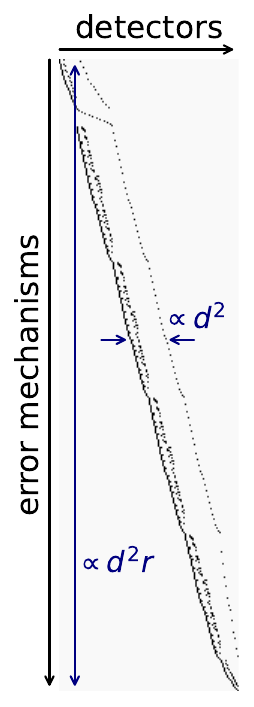}}}
\end{align}
where for simplicity we have just plotted the adjacency matrix between error mechanisms (rows) and detectors (columns).
For a bond dimension $\chi$, the cost of applying each MPO per site grows as $\chi^3$.
Taking all of this into account, the cost of decoding each shot is $\mathcal{O}(\chi^3 d^4 r)$.
Note that we have not studied in depth the scaling of the bond dimension $\chi$ necessary to achieve a given approximation for particular values of $d$, $r$, and error probabilities $\vec{p}$.
We leave this analysis for future work.

\subsection{Why should we expect convergence at small \texorpdfstring{$\chi$}{chi}?}
\label{sec:convergence_expected}

The TNML decoding procedure explained in this section leaves an open question: why should we expect the approximate contraction of the tensor network to converge at low bond dimension $\chi$?
In the $p \rightarrow 0$ limit, the MPS being evolved is at all points a product state: the initial state has all error mechanisms inactive (0) and entanglement is never built up, which makes the computation efficient.
An MPS of bond dimension $\chi=1$ represents this evolution exactly.
At low enough values of $p$, the evolution can be easily approximated as a low-order perturbation of the product state evolution.
If we do not transition to large values of $p$, we expect a small bond dimension MPS to represent well the perturbed evolution.

\clearpage

\begin{widetext}
\section{Additional data}
\label{appendix:additional_data}

\begin{figure}[ht]
    \centering
    \includegraphics[width=0.30\linewidth]{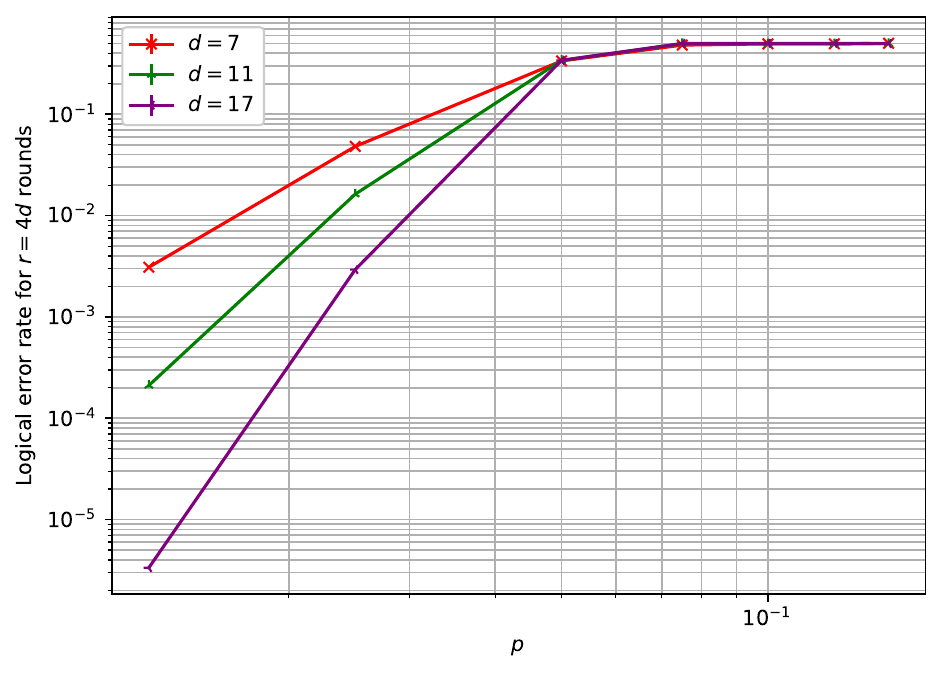}
    \includegraphics[width=0.30\linewidth]{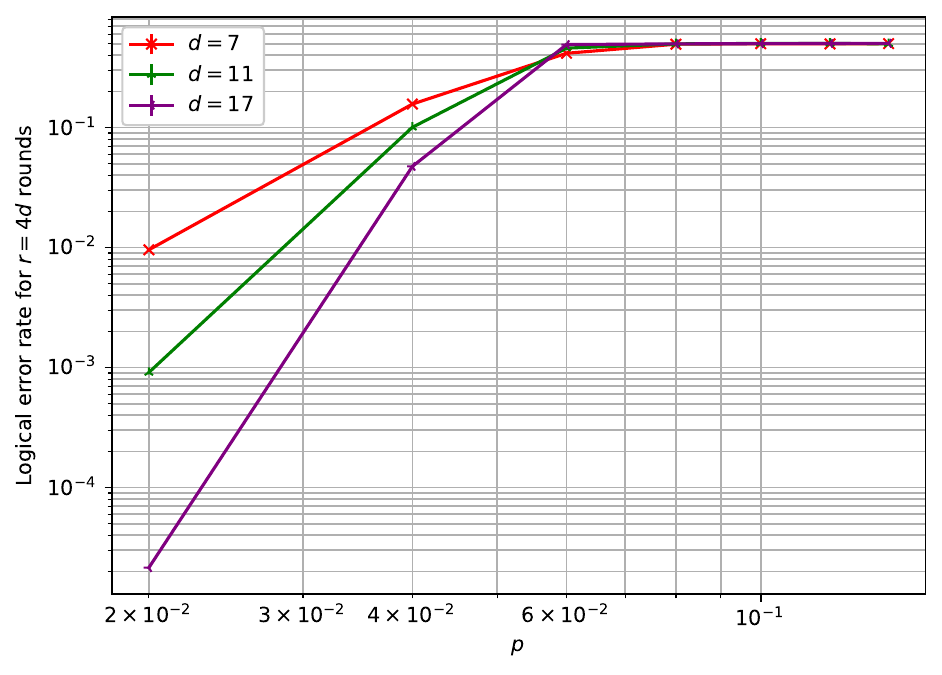}
    \includegraphics[width=0.30\linewidth]{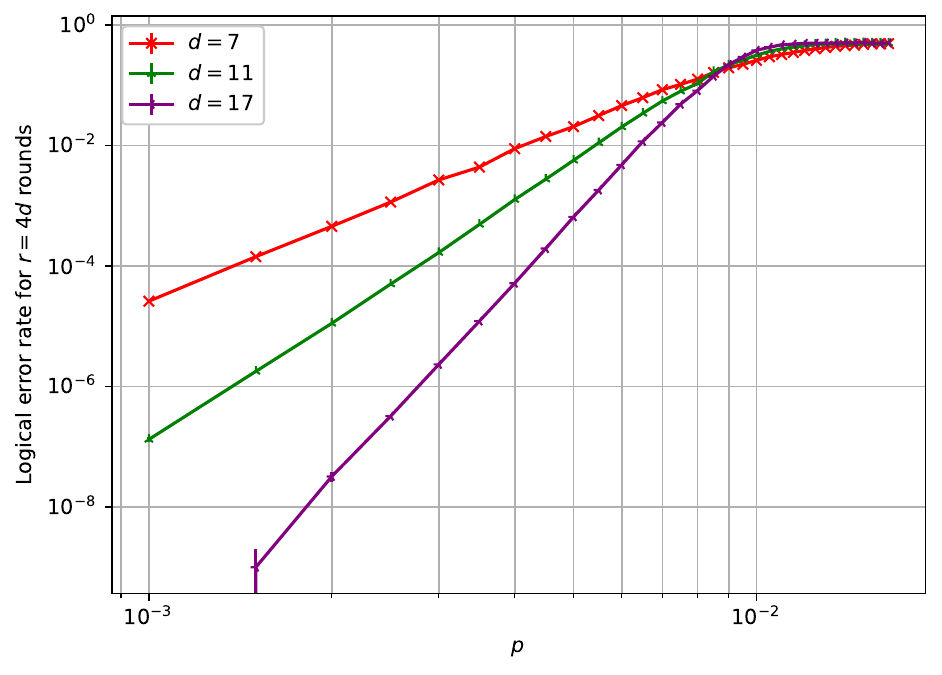}
    \caption{Thresholds of the circuit families considered when decoded using correlated matching. 
    Each shot decoding shot is comprised of $r=4d$ rounds.
    On the left, the repetition code under circuit-level noise has a threshold of approximately $p\approx 0.05$. 
    In the center, the surface code under phenomenological noise has a threshold of approximately $p\approx.058$. 
    On the right, the surface code under circuit-level noise has a threshold of approximately $p\approx.0088$.}
    \label{fig:thresholds}
\end{figure}

\begin{figure}[ht]
    \centering
    \includegraphics[width=\linewidth]{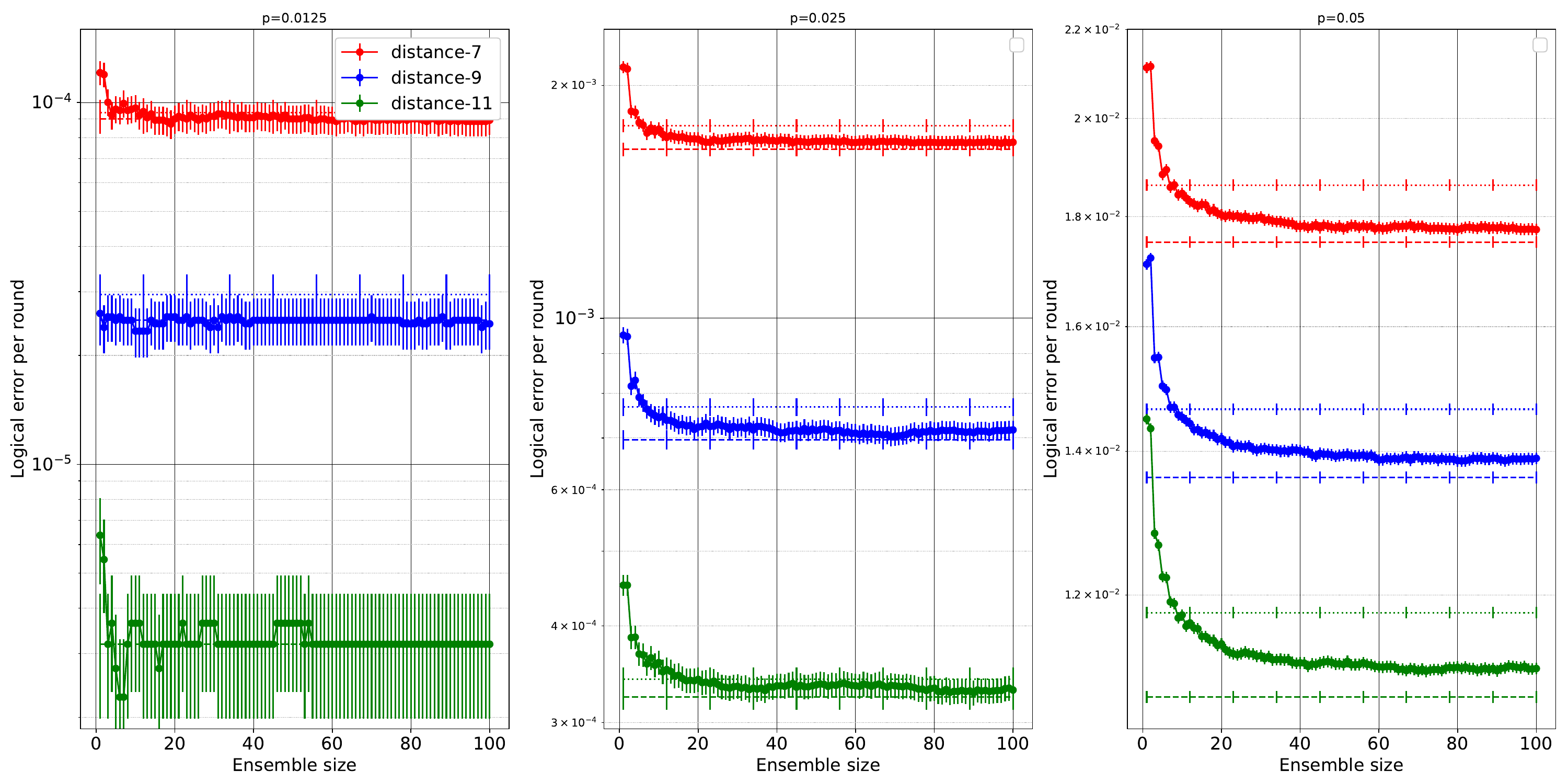}
    \caption{Decoding circuit-level noise in the repetition code at various error rates.
    As in the main text, at each distance there are three lines: minimum-weight matching (dotted), vote pooled ensembled matching (circles), and tensor network decoding (dashed).
    Note that at distance 11 for $p=0.0125$, the minimum-weight matching and tensor network decoder yield the same error rate.
    This is likely due to the small number of failures, and the minimum-weight matching decoder approaching the maximum-likelihood decoder performance at low error rates (as the most likely error dominates).
    }
    \label{rep_code_data_extended}
\end{figure}

\begin{figure}[ht]
    \centering
    \includegraphics[width=\linewidth]{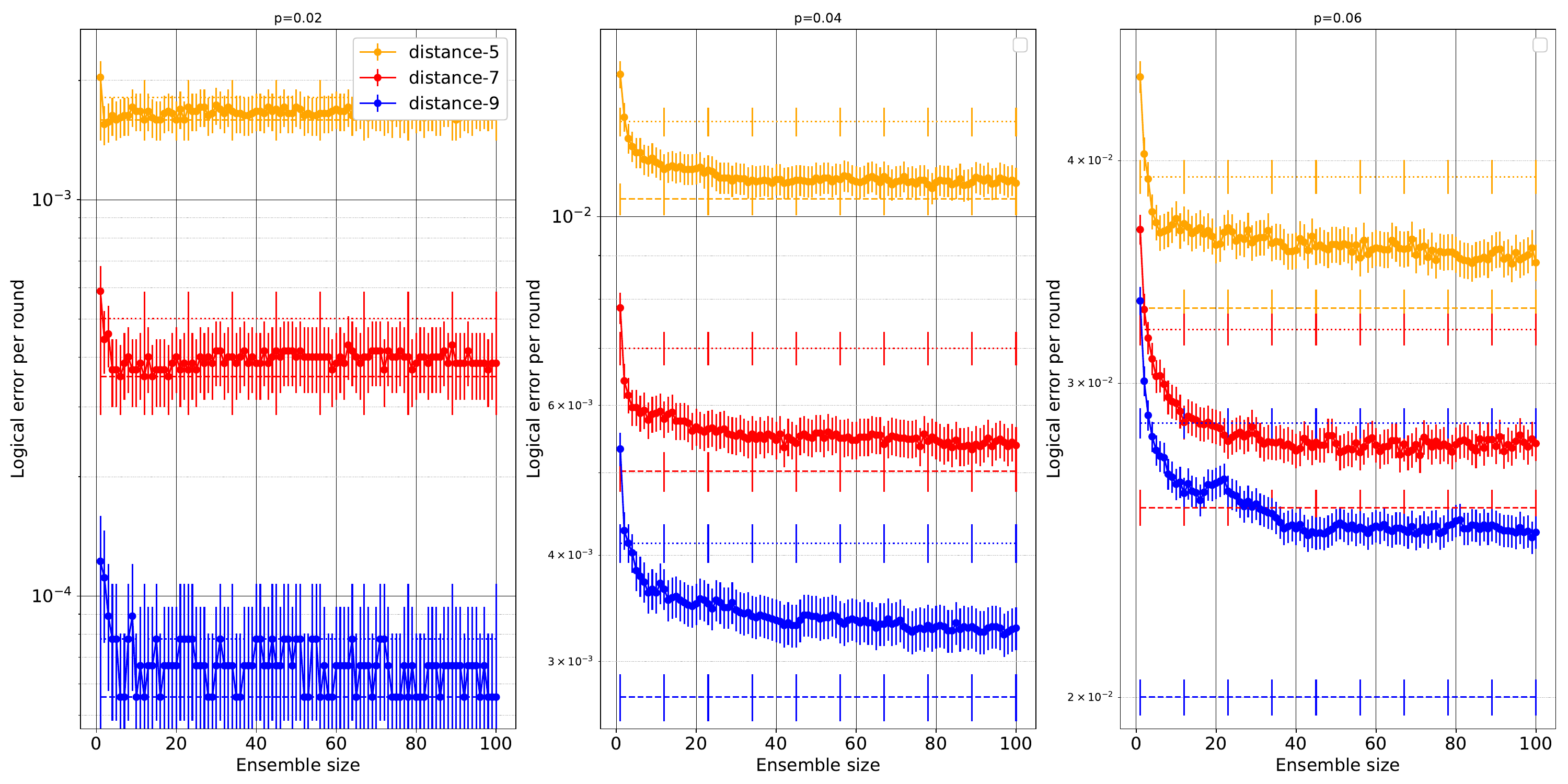}
    \caption{Decoding phenomenological noise in the surface code at various error rates. 
    As in the main text, at each distance there are three lines: correlated matching (dotted), most-likely-error pooled ensembled matching (circles), and TNML decoding (dashed).
    At higher error rates, the gap between ensembled matching and TNML decoding grows larger.
    This illustrates an important point: because the TNML decoder likely boasts a higher threshold, there are error regimes wherein the gap between the TNML decoder and the ensembled matcher will grow.
    However, these very-near-threshold error rates are not the regime in which one would build a quantum memory without overwhelming overhead.
    At more realistic target error rates, the relative performance between the two decoders is similar -- e.g., in the leftmost plot where $p=0.02$, the error rates of ensembled matching and TNML decoding are indiscernable.
    }
    \label{surface_code_data_extended}
\end{figure}

\begin{figure}[ht]
    \centering
    \includegraphics[width=\linewidth]{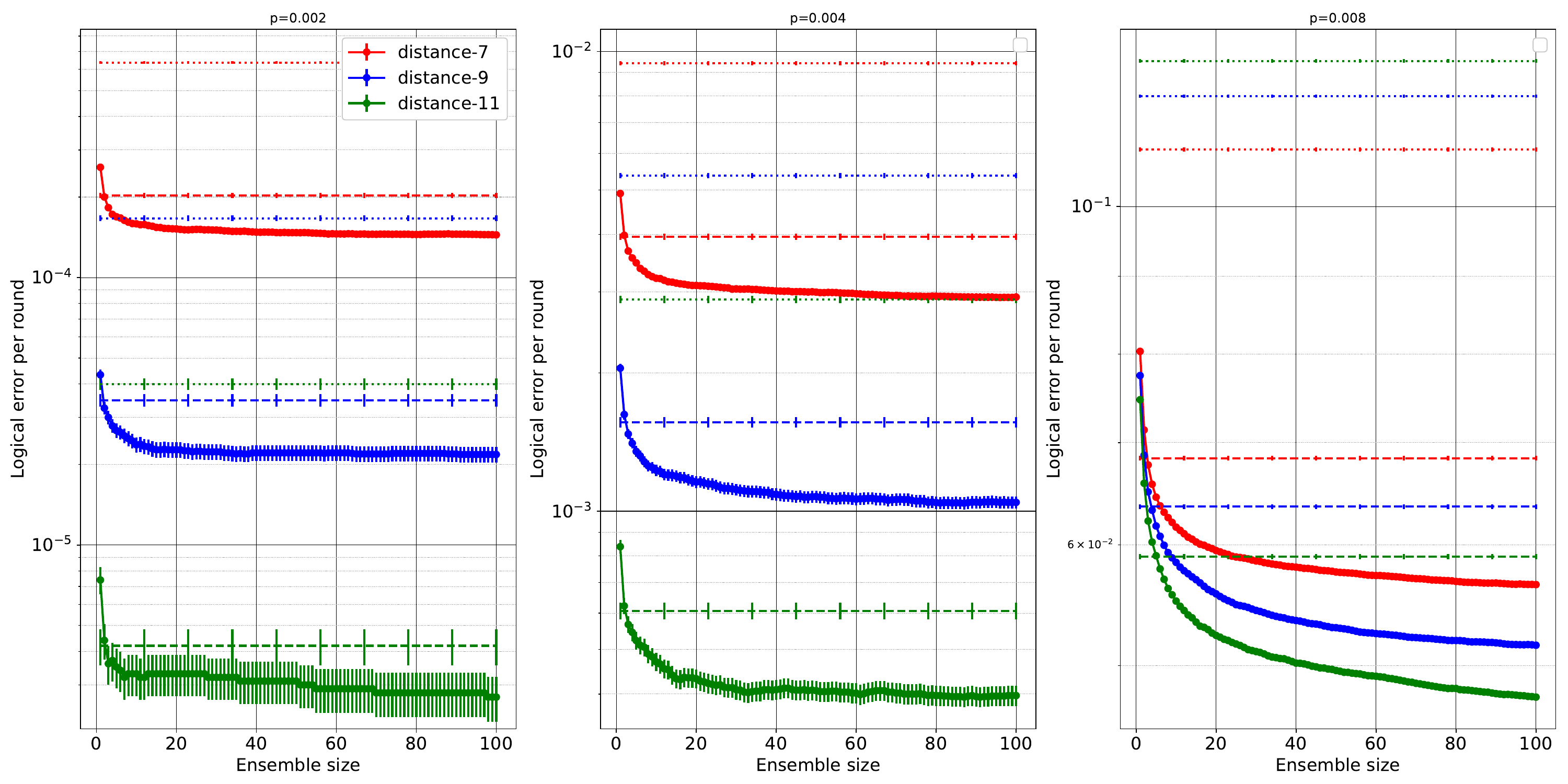}
    \caption{Decoding circuit-level noise in the surface code at various error rates. As in the main text, at each distance there are three lines: minimum-weight matching (dotted), vote pooled ensembled matching (circles), and TNML decoding (dashed). 
    Note the inversion of the lines at $p=0.008$, indicating that this error rate is above the threshold of the uncorrelated decoder but below the threshold of the correlated decoder.
    }
    \label{surface_code_circuit_data_extended}
\end{figure}

\end{widetext}

\end{document}